\documentclass[a4paper, twocolumn]{article}

\usepackage{amssymb}
\usepackage{amstext}
\usepackage{amsmath}
\usepackage{amsthm}
\usepackage{multicol}
\usepackage{url}
\usepackage{graphicx}
\usepackage[square,numbers]{natbib}
\usepackage[margin=2.0cm]{geometry}
\usepackage{appendix}
\usepackage{authblk}

\graphicspath{{figures/}}

\newtheorem{example}{Example}
\newtheorem{definition}{Definition}

\title{Partially Oblivious Neural Network Inference}

\author{Panagiotis Rizomiliotis, Christos Diou, Aikaterini Triakosia, \\
  Ilias Kyrannas and Konstantinos Tserpes}
\affil{\small Department of Informatics and Telematics, \\
  \small Harokopio University of Athens\\
  \small Omirou 9, 17778, Athens, Greece\\
  \small \texttt{\{prizomil,cdiou,ktriakos,kyrannas,tserpes\}@hua.gr}}
\date{}

\begin{document}
\maketitle

\begin{abstract}
  Oblivious inference is the task of outsourcing a ML model, like neural-networks, without disclosing critical and sensitive information, like the  model's parameters. One of the most prominent solutions for secure oblivious inference is based on a powerful cryptographic tools, like Homomorphic Encryption (HE) and/or multi-party computation (MPC).
Even though  the implementation of oblivious inference systems schemes has impressively improved the last decade, there are still significant limitations on the ML models that they can practically implement.  Especially when both the ML model and the input data's confidentiality must be protected. 
In this paper, we introduce the notion of partially oblivious inference. 
We empirically show that for neural network models, like CNNs, some information leakage can be acceptable. We therefore propose a novel trade-off between security and efficiency.
In our research, we investigate the impact on security and inference runtime performance from the  CNN model's weights partial leakage.  We experimentally demonstrate that in a CIFAR-10 network  we can leak up to  $80\%$ of the model's weights  with practically no security impact, while the necessary HE-mutliplications are performed four times faster.
\end{abstract}

\section{Introduction}

Artificial intelligence (AI), and in particular, machine learning (ML) technology is transforming almost every business in the world. ML provides the ability to obtain deep insights from data sets, and to create models that outperform   any human or expert system in critical tasks, like face recognition, medical diagnosis and financial predictions. Many companies offer such ML-based operations as a service  (Machine learning as a service, MLaaS). 
MLaaS facilitates clients to benefit from ML models without the cost of establishing and maintaining an inhouse ML system. There are three parties involved in the transaction; the data owner, the model owner and the infrastructure provider.

However, the use of ML models raises crucial security and privacy concerns. The data set used for the ML model training and/or the MLaaS client's input in the inference phase can leak  sensitive personal or business information. To complete the scenery of security threats, in several applications, like medical or financial, the ML models are considered as the MLaaS provider's intellectual property, and they must be protected.

Oblivious inference is the task of running a ML model without disclosing the client’s input, the model's prediction and/or by protecting the ownership of the trained model. This field of research is also referred to as Privacy-preserved machine learning (PPML).

Several solutions for oblivious inference have been proposed that utilize powerful cryptographic tools, like Multi-party Computation (MPC) primitives and the Homomorphic Encryption (HE) schemes.  MPC based protocols facilitate the computation of an arbitrary function on private input from multiple parties. These protocols have significant communication overhead as they are very interactive. On the other hand, using HE cryptography we are able to perform computations on encrypted data, but with significant computation and storage overhead.

Several PPML schemes have been proposed that are either based solely on one of these technologies or that they leverage a combination of them (hybrid schemes). 
So far, literature has focused on two attack models. It is assumed that  either the model owner is also the infrastructure provider or that the ML model that it is used, it is publicly known. This is a reasonable choice, as in both cases, the ML model's weights can be used in  plaintext form. That is that,  the schemes designers  avoid expensive computations between ciphertexts and thus, they  introduce inference  systems that are practical.

In this paper, we consider the use cases in which the ML model's confidentiality must be protected. 
The service provider wants to  outsource the ML prediction computation (for instance to a cloud provider or to an edge device).  However,  the ML model constitutes intellectual property and it's privacy must be preserved. 

Protecting both the client's input data and the model's privacy can increase prohibitively the computational complexity  as all the computations must be performed between encrypted data.   
Just as a rough estimation, the runtime of a single HE multiplication increases  ten times when it is performed between two ciphertexts compared to HE multiplication between a plaintext and a ciphertext.
At the same time, HE multiplications between encrypted data (ciphertexts)  increase significantly the accumulative level of noise and they limit the applicability of the HE schemes.  
Thus, they must be avoided when possible.

Building on this observation,    we  introduce the notion of {\it partially oblivious (PO) inference}.
In a PO inference system, the ML model owner decides to leak some of the model's weights in order
to improve the efficiency of the inference process. PO inference can be seen as a generalization of oblivious inference that offers a trade-off between security and efficiency. The PO inference systems lie between the two extreme use cases, the most secure but the less efficient in which all the ML model weights are encrypted and the less secure and the most efficient in which all the weights are revealed. 
The optimal point of equilibrium between efficiency and security depends on the use case.

Our work is summarized as follows: 
\begin{enumerate}
\item We introduce the notion of Partially Oblivious inference for ML models.
\item We provide a security definition for the evaluation of the information leakage impact. In our analysis, the attacker is passive ("honest-but-curious") and she aims to compute a model that simulates the protected one as accurately as possible. We use accuracy improvement as our security metric.
\item As a proof-of-concept use case, we apply the notion of PO inference to protect Convolutional Neural Networks (CNN) inference.  
\item We experimentally measure the security and performance trade-off.  We use two models trained with the MINST \cite{mnist} and CIFAR-10 datasets \cite{Krizhevsky09}, respectively. 
For the PO inference  implementation, Gazelle-like \cite{DBLP:conf/uss/JuvekarVC18} approach is used. Impressively, it is shown that in some scenarios, leakage of more than $80\%$ of the model weight's can be acceptable.
\end{enumerate}

The paper is organized as follows. In Section~\ref{sec.back}, the necessary background is provided. In Section~\ref{sec.def}, we analyze our motivation, we introduce the security attack model and the security definition for PO inference and we demonstrate the application of the PO inference to CNN models. Finally, in Section~\ref{sec.4}, we implement and evaluate the two CNN models and in Section~\ref{sec.con}, we conclude the paper.

\subsection{Related work}\label{sec.relate}

There are several lines of work for PPML  systems that leverage advanced cryptographic tools, like MPC and HE. 
The most promising solutions are hybrid, and they are using HE to protect the linear  and MPC to protect the non-linear layers.

CryptoNet (\cite{DBLP:conf/icml/Gilad-BachrachD16}) is the first scheme that deploys the HE primitive for PPML on the MNIST benchmark. 
In the same research line, CHET \cite{DBLP:conf/pldi/DathathriS0LLMM19}, 
SEALion \cite{DBLP:journals/corr/abs-1904-12840} and Faster Cryptonets \cite{DBLP:journals/corr/abs-1811-09953} use HE  and  retrained networks.
  There are HE based schemes that use pre-trained networks, like  Chimera \cite{DBLP:journals/jmc/BouraGGJ20} and Pegasus \cite{DBLP:journals/iacr/LuHHMQ20}. In the pre-trained PPML category, we can find several proposals that use only MPC schemes, like ABY3 \cite{DBLP:conf/ccs/MohasselR18}, 
and  XONN \cite{DBLP:conf/uss/RiaziS0LLK19}.

The most promising type  of PPML systems are hybrid, i.e. the proposals that use both MPC and the HE schemes. Hybrid HE-MPC schemes provide an elegant solution for pre-trained networks. The MPC is responsible for the non-linear part (activation function) and HE for the linear transformations (FC and convolutional layers).
 Gazelle \cite{DBLP:conf/uss/JuvekarVC18}  is a state-of-the-art hybrid scheme for CNN prediction and several works have followed, like Delphi \cite{DBLP:journals/iacr/MishraLSZP20}, nGraph-HE \cite{DBLP:conf/cf/BoemerLCW19},  nGraph-HE2 \cite{DBLP:conf/ccs/BoemerCCW19},  and  PlaindML-HE \cite{DBLP:conf/iccd/ChenCVR19}.  
All these schemes assume that either the model owner runs the models locally or that the ML model is publicly known. 

There are several open source HE libraries that implement the operations of a HE schemes and offer higher-level API \cite{viand2021sok} and there is an ongoing effort to standardize APIs
for HE schemes \cite{standard}. 
However, dealing directly with HE-libraries and operations is still a very challenging task for the developers. In order to facilitate developers work, HE compilers have been proposed to offer a high-level abstraction.   There is a nice overview of existing HE-compilers in~\cite{viand2021sok}.

\section{Background}\label{sec.back}

\subsection{Homomorphic Encryption} 

In the last decade, the performance of HE schemes has impressively improved up
to several orders of magnitude thanks to advances in the theory and to more
efficient implementations. However, it is still significantly slower than
plaintext computations, while realizing HE-based computations is complex for the
non-expert.

Modern HE schemes belong into one of two main categories. The schemes that
compute logical gates and thus, they are most efficient for generic
applications, and the schemes that operate nicely on arithmetic or $p$-ary
circuits and thus, they are used for the evaluation of polynomial functions.
The CKKS \cite{cheon2017homomorphic} scheme belongs in the second category. As
it operates to arithmetic circuits on complex and real approximate numbers, CKKS
is suitable for machine learning applications. We are going to use it in our
experiments.

Following the last version of the HE Standard~\cite{standard}, all the schemes
must support the following types of operations: key and parameters management, encryption and decryption operation, HE evaluation of additions and multiplications, and noise management. 

\subsection{HE evaluation operations cost}\label{sec.cost}
 Practically, all the modern HE schemes are based on the hardness of the {\it
   Learning With Errors (LWE)} problem \cite{DBLP:conf/stoc/Regev05} and its
 polynomial ring variant. Depending on the scheme the plaintext, the keys, and
 the ciphertexts are elements of $\mathbf{Z}_q^n$ or $\mathbf{Z}_q [X]/(X^n+1)$,
 i.e. they are either vectors of integers or polynomials with integer
 coefficients.

In order to protect a message $m$ a randomly selected vector (or polynomial) $e$
is selected from a distribution and it is added to produce a noisy version of
$m$. The level $B$ of this added noise must always be between two bounds
$B_{min}$ and $B_{max}$. When $B<B_{min}$, the ciphertext cannot protect the
message, while when $B>B_{max}$, the noise cannot be removed and the correct
message cannot be retrieved anymore.
 
 Thus, it is crucial to manage the level of noise induced by the HE
 operations. It has been demonstrated that the best noise management approach is
 to treat the ciphertext's noise level $B$ as an invariant. That is that, after
 each HE operation, the level of noise must remain close to $B$.
 
In the CKKS~\cite{cheon2017homomorphic} scheme, the ciphertext is a pair of
polynomials $c = (c_0, c_1)$ over a ring of polynomials $ {Z}_q[X]/(X^N+1)$, for
appropriately selected integers $q$ and $N$. The four main evaluation operations
of CKKS scheme are summarized as:


\begin{enumerate}
\item \textbf{Plaintext-Ciphertext Addition} \\ Let $m$ and $m'$ be two
  plaintexts and $c' = (c'_0, c'_1)$ be the encrypted value of $m'$. The output
  of the addition is $c_{output} = (m + c'_0, c'_1)$ and decrypts to $m + m'$
  and the noise level is $B$.
\item \textbf{Ciphertext Addition} \\ Let $c = (c_0, c_1)$ and $c' = (c'_0,
  c'_1)$ be the encrypted values of plaintexts $m$ and $m'$. The output of the
  addition is $c_{output} = (c_0 + c'_0, c_1 + c'_1)$ and it is the ciphertext
  of $m + m'$ (approximately with good accuracy). The level of noise is upper
  bounded by $2B$.
\item \textbf{Plaintext-Ciphertext multiplication} \\ Let $m$ and $m'$ be two
  plaintexts and $c' = (c'_0, c'_1)$ be the encrypted value of $m'$. The output
  $c_{output} = (m \cdot c'_0, m \cdot c'_1)$ decrypts to $m \cdot m'$ and the
  level of noise is $mB$.
\item \textbf{Ciphertext multiplication} \\ Let $c = (c_0, c_1)$ and $c' =
  (c'_0, c'_1)$ be the encrypted values of plaintexts $m$ and $m'$. The output
  of the multiplication is three polynomials, $c_{output} = (c_0 \cdot c'_0, c_0
  \cdot c'_1 + c'_0 \cdot c_1, c_1 \cdot c'_1)$ and the noise level is $B^2$.
\end{enumerate}

It is clear that the ciphertext multiplication is the problematic one. The number of ciphertext polynomials increases linearly (one more polynomial after each multiplication) and the noise level increases exponentially (it becomes $B^{2^L}$, after $L$ consecutive multiplications).
To manage this size and noise increase, two refresh type operations are applied.   
In order to bring the dimension of the output ciphertext back to two, the relinearlization algorithm is used. The resulting ciphertext  $c''_{output}$
is an encryption of the  $m\cdot m'$ and the level of noise is  $B^2$.
For the noise management, an algorithm called rescale (or modulo switching in other HE schemes)
is used.  However, it can be applied only a  limited and predetermined number of times, usually  equal to the multiplicative depth $L$ of the arithmetic circuit.

Both algorithms, rescaling and relinearization, are costly in terms of computational complexity
and both of them are applied after each multiplication between two ciphertexts.
Relinearization has approximately the same computational cost with ciphertext multiplication and an evaluation key is required. The evaluations keys are created by the encryptor and passed to the evaluator.

To summarize, the HE multiplication between cipehrtexts is a very costly  operation in terms of  computational overhead and noise management. Compared to ciphertext multiplication, the other three HE evaluation operations are practically for free.

\begin{figure*}[!t]
\centering
\includegraphics[width=1.0\textwidth]{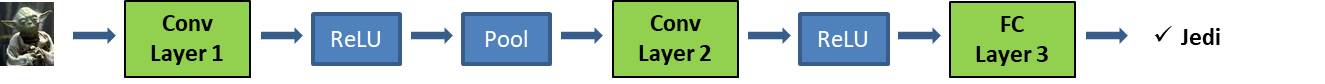}

\caption{The example of a simple CNN. The green blocks are the linear layers (2 convolutional and one fully connected) and blue blocks are the non-linear layers.}
\label{fig_CNN}
\end{figure*}

\subsection{Plaintext Packing}\label{sec.pack}
One of the main features of some HE schemes that extremely improve performance
is plaintext packing (also referred to as batching).  It allows several scalar
values to be encoded in the same plaintext. Thus, for schemes with cyclotomic
polynomial of degree $N$, it is possible to store up to $N/2$ values in the same
plaintext (we refer to them as slots). Thus, homomorphic operations can be
performed component-wise in Single Instruction Multiple Data (SIMD) manner.
This encoding has several limitations, since there is not random access
operation and only cyclic rotations of the slots is allowed.

There are various choices for plaintext packing in ML, i.e. how the input data
and the model weights are organized in plaintexts (or ciphertexts).  Depending
on the workload two are the main packing approaches, {\it batch-axis-packing}
and {\it inter-axis packing}.

The {\it batch-axis-packing} is used by CryptoNets, nGraph-HE and nGraph-HE2. It
is used to a 4D tensor of shape $(B,C,H,W)$, where $B$ is the batch size, $C$ is
number of channels and $H$, $W$ the height and width of input, along the batch
axis. That is that, each plaintext (or ciphertext) packing has $B$ slots and
$C\cdot H\cdot W$ are needed. This approach assumes that $B$ inputs are
available for each inference operation.

On the other hand, {\it inter-axis packing} is used when each input is processed
separately, i.e. it is not necessary to collect $B$ inputs before performing a
prediction (this is common in medical diagnosis). There are several packing
choices, all of them encode scalars from the same input. This approach is used
by Gazelle in which different packing is used for each type of linear
transformation.  We will use {\it inter-axis packing} in our analysis. In
Section~\ref{sec.3}, we provide more details on the different inter-axis packing
choices.

\subsection{CNN models}

The neural-network inference has been identified as the main application area
for privacy preserving technology, and especially for HE and MPC schemes, as we
have seen in Section~\ref{sec.relate}.  However, there are practical limits to
the complexity of the use cases that can be implemented (the unprotected
computation must be at most a few hundreds of milliseconds).

A CNN model consists of linear layers (like convolutional layer and fully
connected layer) and non-linear layers, like an activation function, usually a
ReLU functions or a pooling function, like max-pooling.  A very simple CNN
appears in Fig.~\ref{fig_CNN}.

A fully connected (FC) layer with $M$ inputs and $N$ Outputs is specified by a
tuple $(\mathbf{W},\mathbf{b})$, where $\mathbf{W}$ is an $M\times N$ weight
matrix and $\mathbf{b}$ is a vector of length $N$, called bias. This layer
receives as input vector $\mathbf{v}_{in}$ of length $M$ and computes the output
as the linear transformation of the input: $\mathbf{v}_{out}=
\mathbf{W}^T\mathbf{v}_{in} + \mathbf{b}$.

The convolutional layer has $c_i$ number of input channels with image dimension
$w_i \times h_i$ each and produces $c_o$ output images with dimension $w_o
\times h_o$. The Conv layer is parameterized by $c_i\times c_o$ many $f_w \times f_h$
filters.

\section{Partially Oblivious Inference}\label{sec.def}

\subsection{Attack Model}\label{sec.attmodel}
The architecture of the neural network (number of layers, type of neural
network) is publicly known. On the other hand, the network's weights constitute
intellectual property of MLaaS provider and their confidentiality must be
protected.  We make no assumptions regarding the training data or the training
process. The training can be based on a private dataset as well as on a
partially public one. Also, the ML model may has been trained from scratch or it
may has been based on publicly known pre-trained model.

All the model inference computations are outsourced to a cloud provider or an
edge device (we refer to both as the Server).  We assume that the Server is {\it
  honest-but-curious}, i.e. it executes the operations correctly, but it wants
to reveal any information that it can.  The goal of the attacker is to compute a
ML model that can simulate the original one as accurately as possible.

\begin{figure}[!t]
\centering
\includegraphics[width=0.4\textwidth]{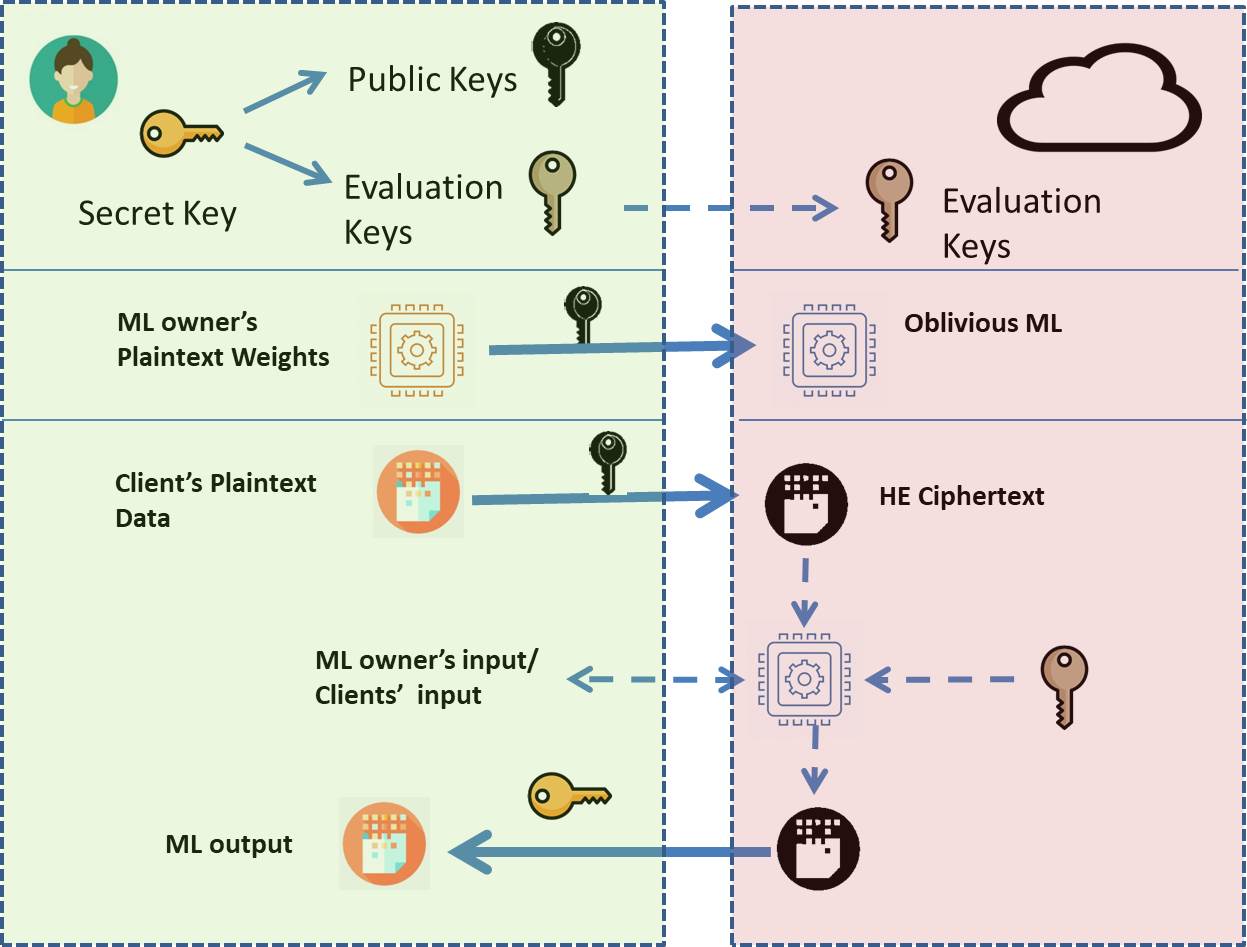}
\caption{Oblivious Inference attack model.}
\label{fig_att_model}
\end{figure}

Our scenario appears in Fig.~\ref{fig_att_model}. We make the assumption that HE schemes are used at least for the linear layers of the ML model. 
In our experiments, we assume that the nonlinear layers are implemented using an MPC protocol (in our case the Garbled Circuit from the Gazelle system). However, our attack model is general  and it  can easily be adapted to other privacy preserving technologies as well.

 The model's owner together with the data owner produce the necessary HE keys, namely, the secret key for the decryption, the public key for the encryption and the corresponding  evaluation keys that are sent to the server. In a simplified version of this scenario, the model's and data owner is the same 
entity. The model's weights and the input data are encrypted with the public key and they are sent to the server.

The server  runs the encrypted model on the encrypted input data, using the evaluation keys for the HE protected linear layers and any other technology for the nonlinear ones. 
The produced output is computed encrypted and it is sent to the legitimate user (it depends on the use case) who can decrypt it using the secret key directly or an MPC protocol on secret key's shares.  

Note that, in theory, we can even hide the architecture of the model, however this is prohibitively expensive and it is avoided in practice.

\subsection{Motivation}\label{sec.mot} 
The efficiency, and practicality, of a HE-based system depends strongly on the type of operations that are performed. Based on the analysis in Section~\ref{sec.cost}, two parameters are crucial:
\begin{enumerate}
\item {\it Multiplicative Depth}: The number of consecutive ciphertext multiplications must remain small. As the depth $L$ increases, the modulo coefficient $q$ must also increase and the HE scheme becomes inefficient.
\item {\it Number of ciphertext multiplications:} Even when multiplications can be performed in parallel, the computational complexity is significant. The goal is to reduce the number of ciphertext multiplications in total. 
\end{enumerate}
 
 In the use cases under consideration the number of ciphertext multiplications is very high
 since both the system's input values and the ML coefficients are HE-encrypted.  
 
 One of the main techniques used to reduce the number of the multiplications between ciphertexts is packing (see Section~\ref{sec.pack}), i.e. to organize several data in the same ciphertext and to compute in parallel all the computations with a single HE ciphertext multiplication.   
Both the human expert and HE-compilers use a pre-processing  phase aiming to optimize the use of packing. For instace CHET,  introduced in \cite{dathathri2019chet}, is  a compiler that  leverages the huge batching capacity and the rotations of the CKKS scheme, to decrease the number of required ciphertext multiplications.

Our goal is to go beyond the capabilities of packing by building on the computational cost asymmetry  between HE ciphertext and HE plaintext multiplications. 
In Table~\ref{tab.oper}, we can see an estimation of the runtime cost for different HE-multiplication related operations, i.e. plaintext multiplication, ciphertext multiplication, rescale and relinearization.
The CKKS-RNS is used implemented in the SEAL library. 

Each entry of the Table~\ref{tab.oper} indicates how many times lower is each operation compared to the plaintext multiplication for $N=2^{12}$ and multiplicative depth $L=2$.
 For instance for the same parameters $N$ and $L$ the ciphertext multiplication is $2.7$ times slower and the relinearization is $16$ times slower. 
 
 While we evaluate each operation separately, in practice the ciphertext multiplication is always followed by the relinearization  operation. That  it that, it is almost $19$ times slower in practice.
Similarly, we can argue for the rescale opeation. 
\begin{table}
\caption{CKKS operations for security level $\lambda=128$-bit. $N$ is the degree of the polynomial ($N/2$  slots per plaintext/ciphertetx) and $L$ is the multiplicative depth. All the parameters are defined in SEAL from the HE standard \cite{standard}. }\label{tab.oper}
\begin{center}
 \begin{tabular}{|c|| c c c|} 
 \hline
  & $N=2^{12}$ & $N=2^{13}$  & $N=2^{14}$  \\   
    & $L=2$ &  $L=4$  &  $L=8$  \\  
 \hline\hline
 Plain Mult. & $1$ & $4$ & $16$ \\ 
 \hline
 Ciph Mult. & $2.7$ & $12.3$ & $60$ \\
 \hline
 Rescale & $7.6$ & $38.5$ & $175.7$ \\
 \hline
 Relinearization & $16.1$ & $80$ & $477$ \\
 \hline
\end{tabular}
\end{center}

\end{table}

From Table~\ref{tab.oper}, it is clear that the multiplications between a plaintext and ciphertext are much more efficient. Based on this observation, we are motivated to investigate the possibility of leaking information that has limited impact on the security of the protected scheme.
The performance benefits are pretty clear in terms of runtime overhead as the revealed values can be used in plaintext form.

\paragraph{Note.} We need to have in mind that plaintext multiplication form leads also to limited noise growth and as a result a larger variety of functions that can be HE implemented.
This is illustrate in Example~\ref{exam.2}. In this paper, we don't investigate this positive side-effect.  
 
\begin{example}\label{exam.2}
Let $f(x,y,z,w)=xy+zw$ be a bilinear function that must be computed homomorphically.
The naive approach requires two ciphertext multiplications, and a single ciphertext addition.

Let's assume that any two of the four input values can be leaked, i.e. they can be used as plaintext.
There are $4 \choose 2$ different  combinations of two inputs, i.e. $6$ in total. Due to the symmetry of the function they belong into two equivalent classes, either both inputs are the operands of the same multiplication (i.e. $\{x,y\}$ or $\{z,w\}$) or operands of different multiplications.

In the first case,  it requires one ciphertext multiplication, and one plaintext-ciphertext addition
(the multiplication between plaintexts is for free).
In the second case,  it requires two plaintext-ciphertext multiplications, and one ciphertext addition. This is much more efficient. 
\end{example}

\subsection{Security Definition}\label{subsec.3}
 In this section, we introduce the notion  of {\it Partially Oblivious (PO)} inference and we provide the corresponding security definition.

Let $\mathcal{W}$ be the set of weights of a model $\mathcal{M}$ and let $\mathcal{L}$ be a
subset of  $\mathcal{W}$.
\begin{definition}
A model $\mathcal{M}$ is $\mathcal{L}$-Partially Oblivious ($\mathcal{L}$-PO), when the model is FHE computed and only the weights $w\not\in \mathcal{L}$ are HE-encrypted.
\end{definition}
The definition implies that the model's inference is computed using HE while the weights that are in the subset $\mathcal{L}$ are used unencrypted, i.e. these weights are leaked.  
 
When $\mathcal{L}=\emptyset$, all the weights are encrypted and 
we have the standard definition of oblivious inference. The model is $\emptyset$-PO.   
 
 Next, we introduce a security definition to assess the impact of the proposed information leakage. 
In our attack scenario, we assume that the adversary's goal is to steal the model. That is that, the attacker wants to produce a model $\mathcal{\hat{M}}$ that is equivalent to $\mathcal{M}$.

Let $\mathcal{\hat{M}}_{\emptyset}$ and $\mathcal{\hat{M}}_{\mathcal{L}}$ be the models that the attacker computes when  all the weights are encrypted and when $\mathcal{L}$ are leaked, respectively. Also, let $ACC_{\emptyset}$ and $ACC_{\mathcal{L}}$  be the accuracy of each model. 

\begin{definition}
A $\mathcal{L}$-PO model $\mathcal{M}$ is   $\lambda$-secure,  if the
advantage of any polynomial-time adversary $A$,
\begin{equation}
Adv_{A}(\mathcal{M},\mathcal{L})= ACC_{\mathcal{L}} - ACC_{\emptyset} 
\end{equation}
is upper bounded by $\lambda$, i.e. 
$$Adv_{A}(\mathcal{M},\mathcal{L})\leq \lambda.$$  
\end{definition}

Ideally, $\lambda=0$. In that case, the leaked weights $\mathcal{L}$
 do not improve the attackers capability to steal the model.

\subsection{CNN model Partially Oblivious Inference}\label{sec.3}
In this section, we investigate the application of our Partially Oblivious  inference approach to the CNN use case. From our attack model, the topology of the network is publicly known, but the weights are confidential.

Our goal is to investigate trade-offs between security and performance. This is expressed as information leakage and more precisely as revealing model's weights. 
We will show that the model's owner can use some of the weights in plaintext in order to improve inference runtime performance, while at the same time this leakage gives a very limited advantage to the attacker.
These weights are only used in the linear layers.

The linear layers (linear operations of fully connected and
convolutional layers) are implemented using an HE scheme. To simplify our
analysis, we assume that the non-linear layers (activation functions) refresh
the ciphertext noise. This is very common in the hybrid schemes. For instance in 
Gazelle \cite{DBLP:conf/uss/JuvekarVC18}, the non-linear layers are implemented with MPC
(using  Garbled Circuits)  and the output of the layer is a ciphertext of with fresh noise (the same applies with TEE based solutions). 

The model owner can reveal a certain percentage of the model's weights. The higher the percentage the more efficient the inference process.  At the same time the security level is decreasing.
The selection of the weights is subject to the restrictions imposed by the packing policy. The weights that are encoded in the same polynomial must be treated as a group. That is that, either they will all be revealed and used in plaintext form or they must all be protected and used in ciphertext form. 

In our investigation, we follow the inter-axis packing (see Section~\ref{sec.pack}) which is more common in medical diagnosis use cases and we analyze different packing techniques for the convolutional and fully connected layers. For the fully connected layer, we analyze the three packing techniques from Gazelle \cite{DBLP:conf/uss/JuvekarVC18}, namely the naive approach, the diagonal and the hybrid. 
On the other hand, the two packing techniques for the convolutional layer (called in the paper padded and packed SISO)  treat each filter value independently. Thus, the model owner can decide on the confidentiality of the convolutional layer weights without restrictions from the packing techniques used. 

The model owner selects  a percentage of the  model weights to reveal following one of the follwoign strategies:
\begin{enumerate}
\item {\it Random selection:} The weight groups of the fully connected layer and the individual filter values of the convolutional layer are selected completely  at random. 
\item{\it Maximum weight:} The weight groups of the fully connected layer and the individual filter weights of the convolutional layer with the largest mean of absolute values are selected.
\end{enumerate}

For example, consider a fully connected layer with $M$ inputs and $N$
outputs. This layer is represented by an $M\times N$ weight matrix
$\mathbf{W}$. The linear output (logit) is $\mathbf{z} =
\mathbf{W}^T\mathbf{h}_{in} + \mathbf{b}$, where $\mathbf{h}_{in}$ is the input
vector and $\mathbf{b}$ the biases. The naive approach with rows as groups will
select $\lfloor pM \rfloor$ rows of $\mathbf{W}$ for encryption. The first
policy will randomly select the rows, while the second will select rows $i$
with the highest mean $\frac{1}{M\cdot N}\sum_{j}\lvert w_{ij}\rvert$.

In the case of a convolutional layer, the layer is represented by a $k \times k
\times M \times N$ tensor $\mathsf{W}$, where $k$ is the kernel size, $M$ is
the number of input channels and $N$ is the number of output channels. In this
case, a group can be a ``filter'' $\mathsf{W}_i$ with dimensions $k \times k
\times M$, or each $k\times k$ kernel $\mathsf{W}_{ij}$. For the maximum weight
strategy, the filters with the largest mean of absolute weights are selected.

For both FC and convolutional layers, encrypting the biases $\mathbf{b}$ is an
additional option (the biases are also encrypted as a group).

Regarding the attacker, we assume that she is trying to produce a model from
the leaked information. We assume that the attacker has a very small set of
data just for the evaluation, but not sufficient data to train or fine-tune a
model. The attacker follows one of the following policies for the prediction of
the missing weights.
\begin{enumerate}
\item {\it Constant ($0.0$):} All the weights are replaced by the constant value (zero,
  in the case of our experiments).
\item {\it Mean ($\mu$):} The mean value of the known weights of the same layer is
  used. If no weights of the current layer are known, then the constant
  policy is used for that layer. 
\item {\it Normal ($\mathcal{N}(0, 1)$):} The values are sampled from a standard normal distribution.
\item {\it Fitted normal ($\mathcal{N}(\mu, \sigma)$):} Same as the
  \emph{Normal} policy, but the values are sampled from a normal distribution
  that is estimated from the known weights of the same layer (i.e., with mean
  equal to the known weight mean, and standard deviation equal to the unbiased
  estimate of the standard deviation of the known weights).
\end{enumerate}

In the following section, we empirically evaluate the tradeoffs between
computational efficiency and security of the proposed $\mathcal{L}$-PO
inference.

\section{Experiments}\label{sec.4}

\subsection{Experimental setup}

For our experiments we have used the MNIST \cite{mnist} and CIFAR-10
\cite{Krizhevsky09} data sets, using the standard train/test splits. The
architecture of the networks used in each dataset are outlined in Table
\ref{tab:nn_arch}.

\begin{table}
  \caption{Neural network architectures used in the experiments with MNIST and
    CIFAR-10 datasets. ``conv\_$\mathbf{3\times 3}$\_$\mathbf{X}$'' denotes a
    convolutional layer with $\mathbf{X}$ filters and a $\mathbf{3\times 3}$
    kernel. FC-$\mathbf{Y}$ is a fully connected layer with $Y$ neurons, while
    ``batchnorm'' denotes batch normalization \cite{ioffe2015batch}. The ReLU
    activation function is used in all layers except the output FC layer.}
  \label{tab:nn_arch}
  \begin{tabular}{|c|c|}
    \hline
    model for MNIST & model for CIFAR-10 \\
    \hline
    Input: ($28\times 28 \times 1$) & Input: ($32\times 32\times 3$)\\
    \hline
    conv\_$3\times 3$\_$32$  & conv\_$3\times 3$\_$32$ \\
    maxpool\_$2\times 2$     & batchnorm               \\
    conv\_$3\times 3$\_$64$  & conv\_$3\times 3$\_$32$ \\
    maxpool\_$2\times 2$     & batchnorm               \\
    FC-10                    & maxpool\_$2\times 2$    \\
    softmax                  & conv\_$3\times 3$\_$64$ \\
                             & batchnorm               \\
                             & conv\_$3\times 3$\_$64$ \\
                             & batchnorm               \\
                             & maxpool\_$2\times 2$    \\
                             & conv\_$3\times 3$\_$128$ \\
                             & batchnorm               \\
                             & conv\_$3\times 3$\_$128$ \\
                             & batchnorm               \\
                             & maxpool\_$2\times 2$    \\
                             & FC-128                  \\
                             & Dropout ($p=0.5$)       \\
                             & FC-10                   \\
                             & softmax                 \\
    \hline
  \end{tabular}
\end{table}

These architectures are used for empirical evaluation of the security and
computational efficiency of the different $\mathcal{L}$-PO strategies.

\subsection{$\mathcal{L}$-PO security}\label{sec.po}

To evaluate the security of $\mathcal{L}$-PO for CNN networks we initially
train both network architectures of Table \ref{tab:nn_arch} using the Adam
optimizer \cite{kingma2014adam} with a learning rate of $0.001$ and categorical
cross-entropy loss. The MNIST network is trained for 10 epochs without the use
of a validation set, while the CIFAR-10 network is trained for 20 epochs, with
20\% validation set and early stopping if no reduction in loss is observed for
more than 3 epochs.

Then, we apply an ``encoding'' step, where weights of the neural network are
selected according to the strategies described in Section \ref{sec.3}. In this
step we simulate $\mathcal{L}$-PO by storing the indices of the weights
that would be selected for encryption. Then, we evaluate the accuracy that an
attacker would achieve in a ``decoding'' step, for different policies (also
described in Section \ref{sec.3}).

Both the original model $\mathcal{M}$, the model $\mathcal{\hat{M}}$ that is
estimated by the attacker, as well as the fully encrypted model
$\mathcal{M}_{\emptyset}$ are evaluated in the test set of each dataset and the
resulting accuracies are used to assess the security of $\mathcal{L}$-PO in
terms of $Adv_{A}(\mathcal{M},\mathcal{L})$ defined in Section \ref{subsec.3}.

In each experiment we define the percentage $p$ of the weights to be selected
for encryption, the weight selection strategy, whether to select biases for
encryption, as well as the attacker policy. Since some of the policies applied
by the attacker are stochastic, we repeat each experiment 10 times and report
the average and standard deviation of $Adv_{A}(\mathcal{M},\mathcal{L})$. For
each experiment (i.e., combination of these options), the original network is
trained only once and is used in all 10 iterations. Training a network for each
different experiment (instead of using the same network across all experiments
of the same datasets) helps take into account randomness introduced by model
training (e.g., due to weight initialization).

Results of the experiments are shown in Table \ref{tab:results}, for both
datasets and for some combinations of $p$, weight selection strategies and
attacker prediction policies. For each experiment, the table provides the
average and standard deviation of the attacker advantage observed in the 10
experiment runs. In all experiments, bias weights have been selected for
encryption.

Note also that, after encrypting all weights ($p = 1.0$) the model accuracy is
roughly equal to the class prior (i.e., the accuracy of the random
classifier). This ascertains our knowledge of the model architeture by itself
does not lead to any attacker advantage.

        \begin{table*}[h!]
        \small
        \begin{center}
        \caption{Results of experiments. Values are average and standard deviation across 10
    runs, in $100Adv_{A}(\mathcal{M}, \mathcal{L})$ (a similar table with the
    corresponding accuracy values is provided in the appendix). All runs include hidden
    biases, while the asterisk `*' indicates the model without any hidden
    weights. Columns $const=0.0$, $\mathcal{N}(0, 1)$, $\mathcal{N}(\mu,
    \sigma)$ and $\mu$ correspond to the different weight estimation policies of
    the attacker.}
        \label{tab:results}
        \begin{tabular}{|c|c|c|c|c|c|c|c|c|}
        \hline
        \multicolumn{9}{|c|}{Random weight selection} \\
        \hline
        & \multicolumn{4}{|c|}{MNIST} & \multicolumn{4}{|c|}{CIFAR10} \\
        $p$ & $ const=0.0$ & $\mathcal{N}(0,1)$ & $\mathcal{N}(\mu, \sigma)$ & $\mu$ & $const=0.0$ & $\mathcal{N}(0, 1)$ & $\mathcal{N}(\mu, \sigma)$ & $\mu$ \\
        \hline
        \hline

            * &
            89.2 &
            88.7 &
            87.8 &
            89.2 &
            70.5 &
            70.9 &
            70.5 &
            70.5 \\
        
                0.0 &
                89.1 (0.0) &
                27.4 (19.5) &
                86.6 (0.9) &
                89.1 (0.0) &
                47.9 (0.0) &
                3.1 (2.3) &
                39.9 (12.5) &
                40.6 (0.0)  \\
            
                0.1 &
                88.3 (0.6) &
                0.7 (3.1) &
                82.2 (4.9) &
                88.4 (0.6) &
                33.3 (11.2) &
                0.0 (0.7) &
                11.6 (9.5) &
                18.4 (13.2)  \\
            
                0.2 &
                85.6 (3.1) &
                0.6 (3.0) &
                62.2 (12.0) &
                83.8 (4.6) &
                18.2 (5.7) &
                0.8 (0.7) &
                5.2 (4.1) &
                9.2 (4.5)  \\
            
                0.3 &
                83.8 (2.1) &
                0.7 (1.0) &
                47.0 (13.7) &
                83.1 (3.3) &
                0.4 (1.0) &
                0.7 (0.8) &
                1.2 (1.7) &
                0.5 (1.4)  \\
            
                0.4 &
                78.3 (4.3) &
                0.1 (2.0) &
                29.8 (13.8) &
                73.4 (6.0) &
                0.6 (1.1) &
                0.9 (1.2) &
                -0.2 (1.4) &
                1.2 (1.5)  \\
            
                0.5 &
                72.0 (9.0) &
                -0.8 (3.2) &
                19.1 (7.9) &
                67.9 (7.6) &
                -0.0 (0.5) &
                0.7 (1.5) &
                0.3 (1.9) &
                -0.0 (1.5)  \\
            
                0.6 &
                47.6 (15.3) &
                0.3 (1.8) &
                8.8 (4.5) &
                40.5 (13.4) &
                -0.4 (0.6) &
                0.4 (0.3) &
                -0.3 (0.6) &
                0.0 (0.0)  \\
            
                0.7 &
                28.3 (9.9) &
                0.6 (1.7) &
                -0.0 (4.3) &
                14.3 (3.8) &
                0.0 (0.0) &
                0.7 (1.1) &
                0.1 (1.1) &
                0.0 (0.0)  \\
            
                0.8 &
                9.6 (7.2) &
                -0.8 (2.7) &
                2.0 (3.0) &
                9.6 (6.6) &
                -0.1 (0.4) &
                -0.1 (0.7) &
                -0.1 (0.7) &
                0.0 (0.0)  \\
            
                0.9 &
                4.6 (5.9) &
                0.4 (3.3) &
                -1.2 (2.9) &
                -0.2 (1.5) &
                0.0 (0.0) &
                -0.2 (1.5) &
                0.6 (1.0) &
                0.0 (0.0)  \\
            
                1.0 &
                0.0 (0.0) &
                0.0 (1.8) &
                0.0 (4.5) &
                0.0 (0.0) &
                0.0 (0.0) &
                0.0 (0.7) &
                0.0 (1.3) &
                0.0 (0.0)  \\

        \hline
        \multicolumn{9}{|c|}{Max weight selection} \\
        \hline
        & \multicolumn{4}{|c|}{MNIST} & \multicolumn{4}{|c|}{CIFAR10} \\
        $p$ & $const=0.0$ & $\mathcal{N}(0,1)$ & $\mathcal{N}(\mu, \sigma)$ & $\mu$ & $const=0.0$ & $\mathcal{N}(0, 1)$ & $\mathcal{N}(\mu, \sigma)$ & $\mu$ \\
        \hline
        \hline

            * &
            89.0 &
            87.8 &
            89.3 &
            89.0 &
            69.7 &
            69.7 &
            69.9 &
            69.7 \\
        
                0.0 &
                88.8 (0.0) &
                16.5 (10.9) &
                86.7 (3.8) &
                88.7 (0.0) &
                28.0 (0.0) &
                2.5 (3.6) &
                27.7 (14.2) &
                17.9 (0.0) \\
            
                0.1 &
                75.5 (0.0) &
                0.2 (2.6) &
                71.0 (8.6) &
                79.9 (0.0) &
                22.2 (0.0) &
                0.1 (0.3) &
                13.4 (10.0) &
                19.8 (0.0) \\
            
                0.2 &
                77.9 (0.0) &
                -1.2 (2.6) &
                60.2 (10.0) &
                76.9 (0.0) &
                8.2 (0.0) &
                -0.2 (1.0) &
                3.5 (2.3) &
                8.9 (0.0) \\
            
                0.3 &
                67.4 (0.0) &
                -2.3 (2.6) &
                25.9 (14.7) &
                43.1 (0.0) &
                0.3 (0.0) &
                0.4 (0.4) &
                0.9 (1.8) &
                1.8 (0.0) \\
            
                0.4 &
                52.1 (0.0) &
                -0.5 (3.8) &
                14.3 (4.3) &
                24.1 (0.0) &
                0.0 (0.0) &
                0.0 (1.0) &
                0.2 (0.9) &
                0.6 (0.0) \\
            
                0.5 &
                33.4 (0.0) &
                -2.5 (2.3) &
                3.6 (3.3) &
                3.9 (0.0) &
                0.0 (0.0) &
                0.3 (0.8) &
                -0.2 (0.9) &
                0.0 (0.0) \\
            
                0.6 &
                39.4 (0.0) &
                -0.7 (2.9) &
                1.6 (2.7) &
                3.5 (0.0) &
                0.0 (0.0) &
                -0.2 (0.8) &
                0.4 (0.5) &
                1.4 (0.0) \\
            
                0.7 &
                15.9 (0.0) &
                -1.9 (1.2) &
                1.9 (3.1) &
                2.5 (0.0) &
                0.0 (0.0) &
                0.3 (0.8) &
                0.4 (1.1) &
                0.0 (0.0) \\
            
                0.8 &
                10.3 (0.0) &
                -0.6 (2.7) &
                1.1 (2.7) &
                8.1 (0.0) &
                0.0 (0.0) &
                -0.1 (0.7) &
                0.4 (0.9) &
                0.0 (0.0) \\
            
                0.9 &
                6.2 (0.0) &
                -0.4 (3.3) &
                1.0 (3.2) &
                6.9 (0.0) &
                0.0 (0.0) &
                -0.1 (0.3) &
                0.3 (1.0) &
                0.0 (0.0) \\
            
                1.0 &
                0.0 (0.0) &
                0.0 (2.3) &
                0.0 (2.4) &
                0.0 (0.0) &
                0.0 (0.0) &
                0.0 (0.6) &
                0.0 (1.3) &
                0.0 (0.0) \\
            
\hline
        \end{tabular}
        \end{center}
        \end{table*}

An important observation from these results is that for both datasets and
models, the attacker advantage quickly diminishes as more weights are
encrypted. This is more pronounced in the more complex model of the CIFAR-10
dataset, where for $p > 0.2$, the attacker advantage is zero for both weight
selection strategies. But even for a small number of hidden weights, e.g.,
$p=0.1$, the maximum attacker advantage (obtained with the constant policy) is
only $47.9$, as opposed to $70.5$ if she or he had access to the full model. In
this case, the attacker would achieve approximately $0.58$ accuracy, while with
the full model she would achieve approximately $0.8$. For the max weight
selection strategy the advantage is even lower.

Another interesting observation is that the weight selection strategy plays an
important role especially for the CIFAR-10 dataset. Max weight selection seems
to be the most effective for both datasets.  On the other hand, random filter and random weight selection conveys the minimum information about how weights were selected to an adversary.
It seems that
the optimal weight selection strategy depends on the model and more
sophisticated methods could be explored. This is not further discussed in this
paper and is left as future work. In the ideal case, one should evaluate
different weight selection strategies and use the one that seems to provide the
best results for each model.

Regarding the different attacker weight estimation policies, replacing all
weights with zero leads to very good results for both datasets, while using
weights from a fitted normal distribution seems to be a good policy as
well. On the other hand, the mean and standard normal polices do not seem to
be as effective.

\begin{figure}
  \includegraphics[width=.9\linewidth]{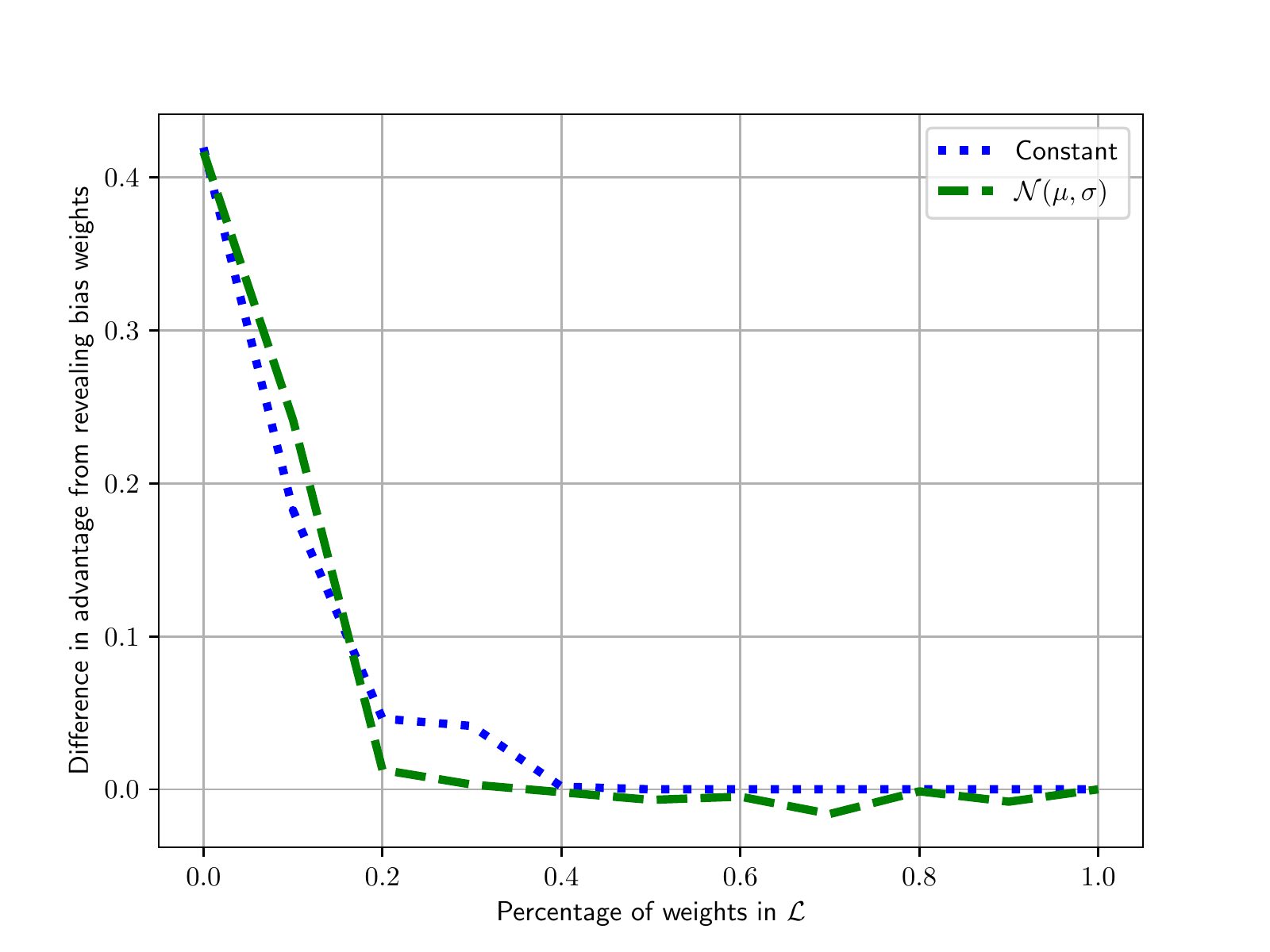}
  \caption{Difference in $Adv_{A}(\mathcal{M},\mathcal{L})$ achieved by an
    attacker by when biases are not included in $\mathcal{L}$, for the max
    weight selection strategy. For very small $p$, the attacker can benefit from
    observing the biases, however as $p$ grows this advantage quickly becomes
    insignificant. }
  \label{fig:bias_advantage}
\end{figure}
In all experiments, the biases have been selected for encryption, since
the addition is relatively cheap, computationally. Figure
\ref{fig:bias_advantage} illustrates the effect of including biases in
$\mathcal{L}$ for the CIFAR dataset for the random and the best two weight estimation policies
for that dataset.

\begin{figure}
  \includegraphics[width=.9\linewidth]{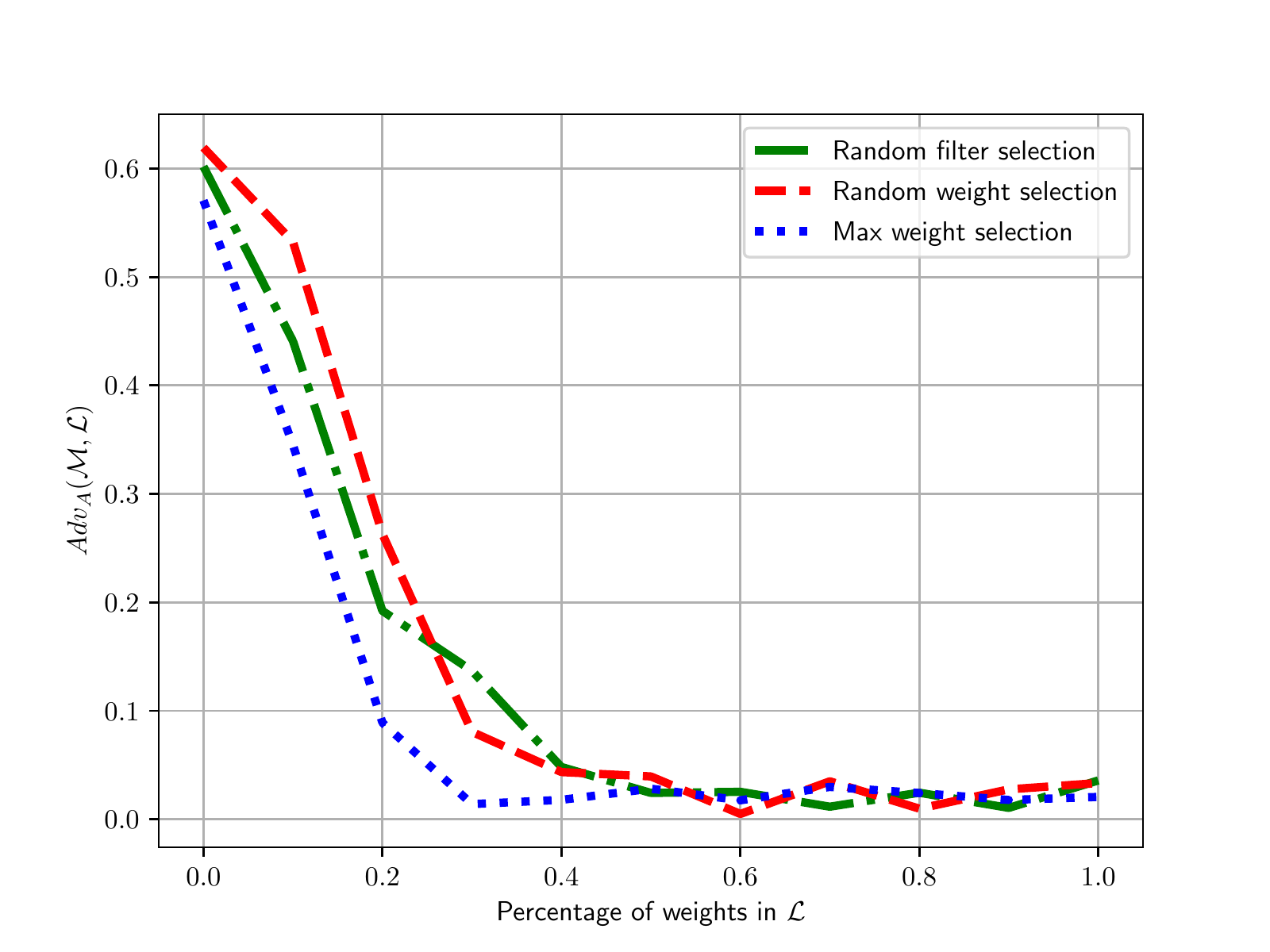}
  \caption{Worst-case across 10 runs, between random filter selection, random
    weight selection and max weight selection on the CIFAR-10 dataset, for
    varying percentages of the weights included in $\mathcal{L}$. In all cases,
    the reference accuracy is approximately  $ACC_{\emptyset}=0.1$, (achieved when setting all
    weights and biases equal to zero). As $p$ grows, the attacker advantage
    quickly diminishes.}
  \label{fig:worst_case}
\end{figure}
Figure \ref{fig:worst_case} illustrates the worst-case in the CIFAR-10
dataset. For each value of $p$, the plot shows the maximum advantage
$Adv_{A}(\mathcal{M},\mathcal{L})$ achieved by the attacker across 10 runs and
across all different weight estimation policies. Ever when encrypting a small
percentage $p$ of the weights, the model effectiveness drops significantly. For
example, the maximum advantage for $p = 0.1$ and the max weight selection
strategy is $0.34$, leading to a model accuracy of approximately $0.44$, which
is significantly worse than the accuracy of the original model (close to
$0.8$). When encrypting more than $20\%$ of the weights, the attacker advantage
becomes insignificant, even in the worst case.

Overall, these results indicate that it is possible to only encrypt a portion
of the weights of a neural network without enabling an attacker to infer the
model weights. In addition, a significant drop in model effectiveness is observed
even when hiding $10\%$ of the weights. Finally, the security was higher for
the more complex CIFAR-10 model, compared to  the simpler MNIST model, especially for
smaller percentages of hidden weights.

\subsection{CNN linear layer micro-benchmarks}
In this section, we evaluate the impact that the selection of the  model weights type (i.e. plaintext or ciphertext) has on the inference performance. We assume that a hybrid inference system is used and more precisely, a system similar to  Gazelle. 

As the model weights are only used in the linear layers, and in order to isolate the impact of the weights leakage to the system's performance, we  have implemented only the linear layers  using the CKKS scheme from SEAL \cite{sealcrypto}.
Regarding the nonlinear layers, we assume that the privacy preserving techniques, used to implement the nonlinear functions,  refresh the noise induced by the HE operations of the linear layer.   
Such an implementation is used in Gazelle that leverages MPC protocols like Garbled Circuits.
Thus, we can further isolate the impact of the leakage selection, as 
 the multiplicative depth per linear layer is very small (usually $L=1$ or $L=2)$ and
we avoid expensive noise refreshing operations. 

The linear layer weights (convolutional and fully connected) are homomorphicaly encrypted
and different packing techniques can be used. 
For our analysis we use the packing techniques from Gazelle.   
For more details on these techniques please refer to the original paper~\cite{DBLP:conf/uss/JuvekarVC18}.

All these techniques mainly perform  ciphertext and/or plaintext  HE multiplications between the layer input and the model weights. The resulting products are always ciphertexts. Depending on the packing technique,  some additions and rotations on these ciphertxts are needed to produce the layer's output. However, these last operations (additions and rotations) and their runtime overhead doesn't depend on the initial type of the model weights, as they always operate on ciphertexts. That is that, only the initial HE multiplications  performance reflects the weight's leakage impact. Thus, we will only consider these multiplications in our experiments.

For all both the FC and the convolutional layers, the input of length  is encrypted in a single ciphertext.
In the FC layer with $M$ inputs and $N$ outputs, the matrix $\mathbf{W}$ is partially encrypted up to percentage $p$. Depending on the packing used, the entries of $\mathbf{W}$ are grouped together in cipheretexts or plaintexts.
In the naive packing, each row constitutes a different group of $M$ elements and there are $N$ groups. Thus, we need $N\cdot (1-p)$ plaintext multiplications and  $N\cdot p$ ciphertext multiplications.  In the diagonal packing, the number of multiplications depends on the inputs $M$.
Similarly, we need   $M\cdot (p-1)$ plaintext and $M\cdot p$ ciphertext multiplications. Finally, the hybrid packing needs the same number of multiplications as the naive techniques. 

We evaluate the overhead of using a partially encrypted matrix $\mathbf{W}$ in Table~\ref{tab.fc.1}. The overhead is computed a multiplicative factor compared to the all weights in plaintext form case. Since after each linear layer the HE noise is refreshed, we avoid the rescale operation. Each ciphertext multiplication is followed by a relinearization operation.

In Table~\ref{tab.fc.2}, we investigate the impact of the rescale operation. Since this operation is used after both the plaintext and the ciphertext operations, the relative performance gain from the weights leakage is reduced.

Finaly, in the convolutional layer, each entry of the $3\times 3$ filters are stored in a different ciphertext and almost all the slots are filled with the corresponding entry's value. Thus,  for each filter application we need $9$ ciphertext or $9$ plaintext multiplications. That is that, we assume that the whole filter is either encrypted or in plaintext form.   
Each convolutional layer input is and RGB image $32\times 32\times 3$.
The result appears in Table~\ref{tab.conv}.

In our experiments we use the convolutional and fully connected layers computed in Section~\ref{sec.po} and we compute the performance of the linear layers for different values of the weights
leakage percentage $p$.

\begin{table}
  \caption{Naive and diagonal packing for the FC layer. The  multiplicative factor of runtime overhead compared to the  all weights in plaintext computation.}\label{tab.fc.1}
  \begin{center}
    \small
    \begin{tabular}{|c|| c c ||c c||}
      \hline
      $p$ &  $(128,10)$  & $(2048,128)$ &  $(128,10)$  & $(2048,128)$  \\
      \hline\hline
      $0.0$ &  $1$ & $1$ &  $1$ & $1$ \\ 
      \hline
      $0.1$ &  $2.9$ & $4.4$ &  $2.8$ & $4.7$\\ 
      \hline
      $0.2$ &  $4.8$ & $7.9$ &  $4.7$ & $8$\\ 
      \hline
      $0.3$ &  $6.7$ & $11.6$ &  $6.6$ & $11.8$\\ 
      \hline
      $0.4$ &  $8.7$ & $15.1$ &  $8.8$ & $15.3$\\ 
      \hline
      $0.5$ &  $10.4$ & $18.3$ &  $10$ & $18.5$\\ 
      \hline
      $0.6$ &  $12.5$ & $22.1$ &  $12.3$ & $22$\\ 
      \hline
      $0.7$ &  $14.3$ & $25.3$ &  $14.2$ & $25.5$\\ 
      \hline
      $0.8$ &  $15.9$ & $29.2$ &  $16$ & $30$\\ 
      \hline
      $0.9$ &  $18.2$ & $32.5$ &  $18$ & $32.8$\\ 
      \hline
      $1.0$ &  $19.8$ & $35.7$ &  $20.1$ & $36.2$\\ 
      \hline
      \hline
    \end{tabular}
  \end{center}
\end{table}

\begin{table}
\caption{Naive  packing for the FC layer using the rescale operation after the plaintext multiplication and using both the rescale and relinearization after the ciphertext multiplication. The  multiplicative factor of runtime overhead compared to the  all weights in plaintext computation.}\label{tab.fc.2}
\begin{center}
 \begin{tabular}{|c|| c c ||} 
 \hline
 $p$ &  $(128,10)$  & $(2048,128)$ \\
 \hline\hline
 $0.0$ &  $1$ & $1$ \\ 
 \hline
 $0.1$ &  $1.3$ & $4.4$ \\ 
 \hline
 $0.2$ &  $1.55$ & $7.9$ \\ 
 \hline
 $0.3$ &  $1.7$ & $11.6$ \\ 
 \hline
 $0.4$ &  $1.9$ & $2.2$ \\ 
 \hline
 $0.5$ &  $2.15$ & $2.5$ \\ 
 \hline
 $0.6$ &  $2.3$ & $2.9$ \\ 
 \hline
 $0.7$ &  $2.5$ & $25.3$ \\ 
 \hline
 $0.8$ &  $2.7$ & $3.252$ \\ 
 \hline
 $0.9$ &  $2.95$ & $3.7$ \\ 
 \hline
 $1.0$ &  $3.11$ & $4.1$ \\ 
 \hline
 \hline
\end{tabular}
\end{center}
\end{table}

\begin{table}
\caption{The convolutional layer using the relinearization operation after the ciphertext multiplication.   All the $9$ multiplications for  a filter application are either all ciphertext or all plaintext. The  multiplicative factor of runtime overhead compared to the  all weights in plaintext computation.}\label{tab.conv}
\begin{center}
 \begin{tabular}{|c|| c ||} 
 \hline
 $p$ &  $(3,3)$   \\
 \hline\hline
 $0.0$ &  $1$  \\ 
 \hline
 $0.1$ &  $4.5$ \\ 
 \hline
 $0.2$ &  $8$  \\ 
 \hline
 $0.3$ &  $11.7$  \\ 
 \hline
 $0.4$ &  $15$  \\ 
 \hline
 $0.5$ &  $18.15$ \\ 
 \hline
 $0.6$ &  $22.3$  \\ 
 \hline
 $0.7$ &  $25.5$  \\ 
 \hline
 $0.8$ &  $29.7$  \\ 
 \hline
 $0.9$ &  $32.55$  \\ 
 \hline
 $1.0$ &  $36.11$ \\ 
 \hline
 \hline
\end{tabular}
\end{center}
\end{table}

\section{Conclusions}\label{sec.con}
This paper initiates a new line of research regarding the oblivious outsourcing
of ML models computation. More specifically, we investigate the trade-offs
between security and efficiency, when some information leakage is acceptable.

Our work serves mainly as a proof of concept using CNNs. We have shown that the
model owner of a CIFAR-10 network can reveal $80\%$ of selected model’s weights
in order to reduce the linear layers cost of multiplications by a factor of
$4$.  While similar security-performance trade-offs are very common in applied
cryptography (in searchable symmetric schemes for instance), it is the first
time that such approach is proposed in ML model inference.

Further research will follow. New attack models must be proposed and new more
fine-grained security definitions must be introduced per use case. At the same
time, the efficiency gain per use case must be evaluated both theoretically
(complexity asymptotic) as well as experimentally. Our goal will be to leverage
the results of this research and provide new design guidelines for efficient
HE-compilers.

\section*{Acknowledgements}
This work was supported by the project COLLABS, funded by the European Commission under Grant Agreements  No. 871518. This publication reflects the views only of the authors, and the Commission cannot be held responsible for any use which may
be made of the information contained therein.

\bibliographystyle{plain}
\bibliography{paper}

\clearpage
\onecolumn
\appendix
\appendixpage
\section*{Additional experimental results}

\begin{table*}[htbp]
  \small
  \begin{center}
    \caption{Results of experiments, in terms of accuracy. Values are average and
      standard deviation across 10 runs, in Accuracy $(\%)$. All runs include
      hidden biases, while the asterisk `*' indicates the model without any
      hidden weights. Columns $const=0.0$, $\mathcal{N}(0, 1)$, $\mathcal{N}(\mu,
      \sigma)$ and $\mu$ correspond to the different weight estimation policies
      of an attacker.}
    \label{tab:appendix}
    \begin{tabular}{|c|c|c|c|c|c|c|c|c|}
      \hline
      \multicolumn{9}{|c|}{Random weight selection} \\
      \hline
      & \multicolumn{4}{|c|}{MNIST} & \multicolumn{4}{|c|}{CIFAR10} \\
      $p$ & $ const=0.0$ & $\mathcal{N}(0,1)$ & $\mathcal{N}(\mu, \sigma)$ & $\mu$ & $const=0.0$ & $\mathcal{N}(0, 1)$ & $\mathcal{N}(\mu, \sigma)$ & $\mu$ \\
      \hline
      \hline      
      * & \multicolumn{4}{|c|}{99.0} &
      \multicolumn{4}{|c|}{80.5} \\
      \hline      
      0.0 &
      98.9 (0.0) &
      37.7 (19.5) &
      97.8 (0.9) &
      98.9 (0.0) &
      57.9 (0.0) &
      12.7 (2.3) &
      49.9 (12.5) &
      50.6 (0.0)  \\
      
      0.1 &
      98.1 (0.6) &
      11.0 (3.1) &
      93.3 (4.9) &
      98.2 (0.6) &
      43.3 (11.2) &
      9.6 (0.7) &
      21.5 (9.5) &
      28.4 (13.2)  \\
      
      0.2 &
      95.4 (3.1) &
      10.8 (3.0) &
      73.3 (12.0) &
      93.6 (4.6) &
      28.2 (5.7) &
      10.3 (0.7) &
      15.2 (4.1) &
      19.1 (4.5)  \\
      
      0.3 &
      93.6 (2.1) &
      11.0 (1.0) &
      58.1 (13.7) &
      92.9 (3.3) &
      10.4 (1.0) &
      10.3 (0.8) &
      11.2 (1.7) &
      10.5 (1.4)  \\
      
      0.4 &
      88.1 (4.3) &
      10.3 (2.0) &
      41.0 (13.8) &
      83.2 (6.0) &
      10.6 (1.1) &
      10.4 (1.2) &
      9.8 (1.4) &
      11.2 (1.5)  \\
      
      0.5 &
      81.8 (9.0) &
      9.5 (3.2) &
      30.3 (7.9) &
      77.7 (7.6) &
      10.0 (0.5) &
      10.2 (1.5) &
      10.2 (1.9) &
      10.0 (1.5)  \\
      
      0.6 &
      57.4 (15.3) &
      10.6 (1.8) &
      19.9 (4.5) &
      50.3 (13.4) &
      9.7 (0.6) &
      10.0 (0.3) &
      9.6 (0.6) &
      10.0 (0.0)  \\
      
      0.7 &
      38.1 (9.9) &
      10.8 (1.7) &
      11.1 (4.3) &
      24.1 (3.8) &
      10.0 (0.0) &
      10.2 (1.1) &
      10.1 (1.1) &
      10.0 (0.0)  \\
      
      0.8 &
      19.4 (7.2) &
      9.5 (2.7) &
      13.2 (3.0) &
      19.4 (6.6) &
      9.9 (0.4) &
      9.4 (0.7) &
      9.8 (0.7) &
      10.0 (0.0)  \\
      
      0.9 &
      14.4 (5.9) &
      10.7 (3.3) &
      10.0 (2.9) &
      9.6 (1.5) &
      10.0 (0.0) &
      9.3 (1.5) &
      10.6 (1.0) &
      10.0 (0.0)  \\
      
      1.0 &
      9.8 (0.0) &
      10.3 (1.8) &
      11.1 (4.5) &
      9.8 (0.0) &
      10.0 (0.0) &
      9.6 (0.7) &
      9.9 (1.3) &
      10.0 (0.0)  \\

      \hline
      \multicolumn{9}{|c|}{Max weight selection} \\
      \hline
      & \multicolumn{4}{|c|}{MNIST} & \multicolumn{4}{|c|}{CIFAR10} \\
      $p$ & $const=0.0$ & $\mathcal{N}(0,1)$ & $\mathcal{N}(\mu, \sigma)$ & $\mu$ & $const=0.0$ & $\mathcal{N}(0, 1)$ & $\mathcal{N}(\mu, \sigma)$ & $\mu$ \\
      \hline
      \hline

      * & \multicolumn{4}{|c|}{98.8} &
      \multicolumn{4}{|c|}{79.7} \\
      \hline
      
      0.0 &
      98.6 (0.0) &
      27.6 (10.9) &
      96.2 (3.8) &
      98.5 (0.0) &
      38.0 (0.0) &
      12.5 (3.6) &
      37.5 (14.2) &
      27.9 (0.0) \\
      
      0.1 &
      85.2 (0.0) &
      11.3 (2.6) &
      80.5 (8.6) &
      89.7 (0.0) &
      32.2 (0.0) &
      10.1 (0.3) &
      23.2 (10.0) &
      29.8 (0.0) \\
      
      0.2 &
      87.7 (0.0) &
      9.9 (2.6) &
      69.7 (10.0) &
      86.7 (0.0) &
      18.2 (0.0) &
      9.8 (1.0) &
      13.3 (2.3) &
      18.9 (0.0) \\
      
      0.3 &
      77.2 (0.0) &
      8.8 (2.6) &
      35.4 (14.7) &
      52.9 (0.0) &
      10.3 (0.0) &
      10.4 (0.4) &
      10.7 (1.8) &
      11.8 (0.0) \\
      
      0.4 &
      61.9 (0.0) &
      10.6 (3.8) &
      23.8 (4.3) &
      33.9 (0.0) &
      10.0 (0.0) &
      10.0 (1.0) &
      10.1 (0.9) &
      10.6 (0.0) \\
      
      0.5 &
      43.2 (0.0) &
      8.6 (2.3) &
      13.2 (3.3) &
      13.7 (0.0) &
      10.0 (0.0) &
      10.3 (0.8) &
      9.6 (0.9) &
      10.0 (0.0) \\
      
      0.6 &
      49.2 (0.0) &
      10.4 (2.9) &
      11.1 (2.7) &
      13.3 (0.0) &
      10.0 (0.0) &
      9.8 (0.8) &
      10.2 (0.5) &
      11.4 (0.0) \\
      
      0.7 &
      25.7 (0.0) &
      9.2 (1.2) &
      11.5 (3.1) &
      12.3 (0.0) &
      10.0 (0.0) &
      10.3 (0.8) &
      10.2 (1.1) &
      10.0 (0.0) \\
      
      0.8 &
      20.1 (0.0) &
      10.5 (2.7) &
      10.6 (2.7) &
      17.9 (0.0) &
      10.0 (0.0) &
      10.0 (0.7) &
      10.3 (0.9) &
      10.0 (0.0) \\
      
      0.9 &
      16.1 (0.0) &
      10.7 (3.3) &
      10.5 (3.2) &
      16.7 (0.0) &
      10.0 (0.0) &
      9.9 (0.3) &
      10.2 (1.0) &
      10.0 (0.0) \\
      
      1.0 &
      9.8 (0.0) &
      11.1 (2.3) &
      9.5 (2.4) &
      9.8 (0.0) &
      10.0 (0.0) &
      10.0 (0.6) &
      9.8 (1.3) &
      10.0 (0.0) \\      
      \hline
    \end{tabular}
  \end{center}
\end{table*}

\end{document}